\documentclass[conference, final]{IEEEtran}
\IEEEoverridecommandlockouts
\usepackage{amsmath,amssymb,amsfonts}
\usepackage{algorithmic}
\usepackage{graphicx}
\usepackage{textcomp}
\usepackage{xcolor}
\usepackage{bm}
\usepackage{subfigure}
\usepackage{hyperref}

\def\BibTeX{{\rm B\kern-.05em{\sc i\kern-.025em b}\kern-.08em
    T\kern-.1667em\lower.7ex\hbox{E}\kern-.125emX}}
\begin{document}

\title{Multi-Context Dual Hyper-Prior Neural Image Compression
 \thanks{ This research is based upon work supported by the National Aeronautics and Space Administration
(NASA), via award number 80NSSC21M0322 under the title
of \emph{Adaptive and Scalable Data Compression for Deep Space
Data Transfer Applications using Deep Learning}.}
}

\author{
     Atefeh Khoshkhahtinat$^\dag$, Ali Zafari$^\dag$, Piyush M. Mehta$^\ddag$,\\ Mohammad Akyash$^\dag$, Hossein Kashiani$^\dag$, Nasser M. Nasrabadi$^\dag$\\
    $^\dag$Dept. of Computer Science \& Electrical Engineering, West Virginia University, WV USA\\
    $^\ddag$Dept. of Mechanical \& Aerospace Engineering, West Virginia University, WV USA\\
    {
    \tt\small \{\href{mailto:ak00043@mix.wvu.edu}{ak00043},\href{mailto:az00004@mix.wvu.edu}{az00004}\}@mix.wvu.edu,\href{mailto:piyush.mehta@mail.wvu.edu}{piyush.mehta}@mail.wvu.edu,
    }\\
    {
    \tt\small \{\href{mailto:ma00098@mix.wvu.edu}{ma00098},\href{mailto:hk00014@mix.wvu.edu}{hk00014}\}@mix.wvu.edu,\href{mailto:nasser.nasrabadi@mail.wvu.edu}{nasser.nasrabadi}@mail.wvu.edu
    
    }
}

\maketitle

\begin{abstract}

Transform and entropy models are the two core components in deep image compression neural networks. Most existing learning-based image compression methods utilize convolutional-based transform, which lacks the ability to model long-range dependencies, primarily due to the limited receptive field of the convolution operation. To address this limitation, we propose a Transformer-based nonlinear transform. This transform has the remarkable ability to efficiently capture both local and global information from the input image, leading to a more decorrelated latent representation. In addition, we introduce a novel entropy model that incorporates two different hyperpriors to model cross-channel and spatial dependencies of the latent representation. To further improve the entropy model, we add a global context that leverages distant relationships to predict the current latent more accurately. This global context employs a causal attention mechanism to extract long-range information in a content-dependent manner. Our experiments show that our proposed framework performs better than the state-of-the-art methods in terms of rate-distortion performance.   

\end{abstract}


\section{\textbf{Introduction}}

Neural image compression which is crucial to reduce the storage or transmission capacity has gained significant popularity in the computer vision field. Recently, learning-based image compression methods have achieved notable progress compared with hand-engineered codecs such as JPEG\cite{jpeg} and JPEG2000 \cite{jpeg2000}. Most of the image compression networks are based on the variational autoencoder (VAE) \cite{kingma2013auto} which is comprised of the two key sub-networks. The first network is the main autoencoder, which is employed to map the image into a compact latent representation and transform the representation back into the original image. The second network is the entropy model which is required to estimate the entropy of the latent representation. It is apparent that improving these
two sub-networks will lead to boosting compression performance. Obtaining more decorrelated and compact representation depends on the main autoencoder capability for extracting dependencies. In addition, building a powerful and accurate
entropy model results in deriving less bit-rate.

A majority of learned image compression models employ
a convolutional encoder to exploit image spatial correlations to
achieve compressible latent representation; however, convolutional neural networks (CNNs)
have some drawbacks. First, the convolution operation is only capable of leveraging local spatial dependencies within the local receptive field. Second, for various inputs, the weights of convolution filters are fixed after training. To address these issues, several solutions have been considered. Motivated by the success of attention in computer vision tasks, some works combined non-local attention modules with convolutional layers to extract long-range correlation for better compression efficiency \cite{chen2021end}, \cite{fu2021learned}, \cite{zou2022devil}. However, such attention modules still do not affect the inherent local-aware property of the CNN architecture. With the rapid development of vision Transformers, Transformer-based autoencoders have been applied in image compression tasks \cite{zhu2022transformer}. In \cite{zhu2022transformer}, Swin Transformer \cite{liu2021swin} is used to build the non-linear transform which is further improved in terms of rate-distortion performance by enforcing frequency decomposition on a feature level \cite{zafari2023frequency}. Swin Transformer is designed upon the window-based attention and shifted-window-based attention to tackle the high computational complexity issue of ViT \cite{dosovitskiy2021an}. Despite the progress, the receptive field of the Swin Transformer is not wide enough to capture global information.

Since generating an optimum compressed bitstream relies on the entropy model, learning an accurate entropy model is vital. To this end, various works have been proposed. Ball{\'{e} \emph{et al.} \cite{balle2018a} introduce a hyperprior that captures the spatial dependencies of the latent representation. The entropy model is approximated as a Gaussian scale mixture (GSM) where the
scale parameters are predicted by the decoded hyperprior. Inspired by the PixelCNN \cite{van2016conditional}, previous works \cite{minnen2018joint}, \cite{lee2018contextadaptive} enjoy the autoregressive component in the entropy model, which utilizes adjacent local latent representations that have already been decoded to predict the distribution of the uncoded latent. Qian \emph{et al.}\cite{qian2021learning} present a global reference model to explore long-term connections. This algorithm searches throughout the previously decoded elements to find the most similar latent for predicting the current latent. While these approaches contribute to estimating precisely the probability distribution of the latent representation, they still have limitations in modeling correlations. First, the hyperprior aims solely to exploit spatial dependencies of the latent space. Second, the proposed context models cannot leverage information from all the decoded elements to approximate the distribution parameters of the current element.

To address the aforementioned challenges, we propose a novel learned image compression network. The major contributions of this work can be summarized as follows:  

\begin{enumerate}

\item The main autoencoder is built upon the Transformer termed Aggregated-window Transformer (AGWinT), which benefits from  both local and aggregated-window Transformer blocks. Without increasing the computational complexity, the global information is extracted by using an aggregated-window Transformer block to achieve a more decorrelated latent representation.

\item The global context is incorporated into the entropy model to leverage long-range correlations in the latent space. This type of entropy model provides accurate probability estimation of the latent representation.

\item Two hyperpriors are introduced, one of them attempts to model inter-channel dependencies and the
other one is utilized to extract inter-spatial dependencies among the elements of the latent representation via the attention mechanism.

\end{enumerate}


\label{sec:format}

\section{\textbf{Related Work}}
\subsection{\textbf{Learned Image Compression}}

Nowadays, the cutting-edge approach for lossy image compression primarily involves a combination of variational autoencoders (VAEs) and transform coding techniques \cite{goyal2001theoretical}. This innovative approach goes beyond traditional linear transformations and incorporates learned non-linear transformation blocks. In the learned image compression framework, at the encoder side, the input image is passed through the analysis transform $g_a(\bm{x};\bm{\phi})$ to generate its compact latent representation $\bm{y}$. Then, to reduce the number of bits required to store or transmit the image, the continuous latent embeddings $\bm{y}$ are quantized to discrete symbols $\bm{\hat{y}}$, which will be subjected to lossless coding algorithms to obtain bitstream files. To reconstruct the compressed image $\bm{\hat{x}}$, the decoder retrieves the quantized latent values $\bm{\hat{y}}$ from the encoded bitstream. It then employs the synthesis transform $g_s(\bm{\hat{y}};\bm{\theta})$ , which is designed to closely approximate the inverse of the analysis transform \cite{balle2020nonlinear}. The process can be summarized as follows:
\begin{equation}
 \begin{aligned}
    &\bm{y}= g_a(\bm{x};\bm{\phi})\\ &\bm{\hat{y}}=Q(\bm{y}),
    \\&\bm{\hat{x}}= g_s(\bm{\hat{y}};\bm{\theta}).
\end{aligned}
\end{equation}
Ball{\'{e} \emph{et al.} \cite{balle2016end}  proposed the factorized density model, which is shared between the encoder and decoder, to effectively exploit the remaining coding redundancy within the latent space. This model is designed to estimate the latent distribution by employing local histograms. In a subsequent study, Ball{\'{e} \emph{et al.} \cite{balle2018a} introduced hidden latent variables, represented as $\bm{\hat{z}}$, to serve as side information for capturing spatial dependencies among the elements of the latent representation. This model employs additional analysis and synthesis transforms, represented as $h_a(\bm{\hat{z}};\bm{{\phi}_h})$ and $h_s(\bm{\hat{z}};\bm{{\theta}_h})$ respectively, to conditionally model the distribution  with respect to the side information. The distribution of $\bm{\hat{z}}$ is approximated using a factorized density model and $p_{\bm{\hat{y}|\hat{z}}}(\bm{\hat{y}|\hat{z}})$ is modeled as a zero-mean Gaussian distribution. The entropy model mechanism of this proposed network can be summarized as: 
\begin{equation}
 \begin{aligned}
    &\bm{z}= h_a(\bm{y};\bm{\phi}_h)\\ &\bm{\hat{z}}=Q(\bm{z}),
    \\& P_{\bm{\hat{y}}|\bm{\hat{z}}}(\bm{\hat{y}}|\bm{\hat{z}}){\sim} {\mathcal{N}}(0,\,h_s(\bm{\hat{z}};\bm{\theta}_h)).
\end{aligned}
\end{equation}

\begin{figure*}[tp]
    \centering
    \includegraphics[width=.85\linewidth]{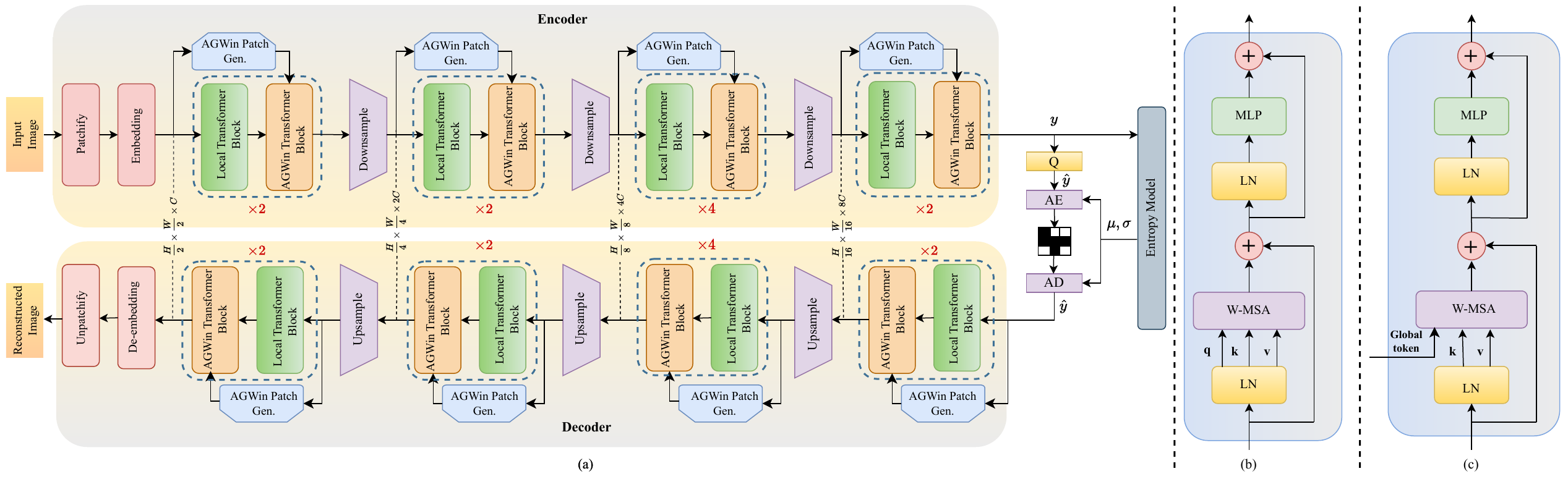}
    \caption{Overview of our proposed AGWinT-based architecture. (b) Local Transformer block. (c) Aggregated-window Transformer block.}
 
    \label{fig:visual-comparison}
\end{figure*}

\subsection{\textbf{Autoregressive-based Entropy Model}}
As the entropy model approaches a closer approximation to the true distribution of the latent representation, the resulting compressed file exhibits a lower bit rate. Minnen \emph{et al.} \cite{minnen2018joint} leveraged the concepts from PixelCNN \cite{van2016conditional} to expand the Gaussian scale mixture-based (GSM) entropy model into a Gaussian mixture model (GMM). This extension involves integrating a context block, which adopts an autoregressive model, along with a hyperprior network. Within this entropy model, the mean and scale parameters of the latent representation distribution are dependent on the hyperprior and causal context associated with each latent $\hat{y}_i$. Therefore, the predicted Gaussian parameters for the distribution of each latent element $\hat{y}_i$ can be expressed as follows:
\begin{equation}
(\mu_i,\sigma_i)=g_{ep}(g_{cm}(\bm{\hat{y}_{<i}};\bm{{\theta}_{cm}}),h_s(\bm{\hat{z}};\bm{{\theta}_h}),\bm{{\theta}_{ep}}),
\end{equation}

The context model $g_{cm}(.)$ is designed by employing a 2D masked convolution. $g_{ep}(.)$ shows the entropy parameter function and $\bm{\hat{y}_{<i}}$ denotes the causal decoded neighbors of current latent element $\hat{y}_i$.

In \cite{zhou2019multi,cui2020g,cui2021asymmetric}, researchers proposed a multi-scale context model that utilized multiple masked convolutions with varying kernel sizes. This allowed the model to learn diverse spatial dependencies simultaneously. On the other hand, \cite{mentzer2018conditional,liu2019practical,chen2019neural} used 3D masked convolutions to leverage both cross-channel correlations and spatial dependencies together. Qian \emph{et al.} \cite{qian2021learning} introduced a global reference context model that aims to exploit long-range dependencies. This component examines previously decoded elements to identify the most similar latent element which is then utilized to estimate the distribution parameters of the current latent. He developed a checkerboard context model which utilizes information from closer neighboring latents to predict the uncoded latent.

\subsection{\textbf{Transformers}}
The Transformer \cite{vaswani2017attention} was originally introduced in the field of natural language processing (NLP) and had a profound impact on the NLP domain. Its remarkable success in NLP applications inspired researchers to extend the Transformer architecture to computer vision (CV) tasks, including object detection\cite{carion2020end}, image classification \cite{touvron2021training}, semantic segmentation \cite{wang2021end}, data compression \cite{zafari2022attention}, and time-series analysis \cite{farahani2023time}. ViT \cite{dosovitskiy2020image} is the first pure Transformer architecture that has demonstrated superior performance compared to convolutional neural network (CNN) models like ResNet \cite{targ2016resnet,mohamadi2023fussl} and attain state-of-the-art performance on multiple image classification benchmarks \cite{sun2017revisiting,mahajan2018exploring,safavigerdini2023predicting,tanhaeean2023analyzing}. 
Vit framework begins by partitioning the input image into non-overlapping patches. These patches are then passed through a Transformer architecture, which leverages self-attention mechanisms \cite{talemi2023aaface} to learn the uniform short and long-range relationships within the image more effectively. However, ViT is effective at capturing contextual information, its monolithic architecture and the quadratic computational complexity associated with self-attention pose significant obstacles in promptly deploying the model for vision tasks, where processes high-resolution images. Several works \cite{liu2021swin,dong2022cswin,tu2022maxvit,Ansari2023An}, particularly the Swin Transformer \cite{liu2021swin}, have been proposed to focus on achieving a balance between short and long-range spatial dependencies through using hierarchical architectures. The Swin Transformer architecture introduces a solution where self-attention is computed within local windows, and a shifted-window self-attention mechanism is employed to model interactions across different windows. Despite significant progress, the self-attention mechanism struggles to capture long-range information due to the constrained receptive field of local windows. Additionally, window-shifted attention only consider small nearby regions around each window when computing interactions. To overcome this limitation, Focal Transformer \cite{yang2021focal} is introduced by implementing more intricate self-attention modules, but this improvement comes at the cost of increased computational complexity.



\section{\textbf{Proposed Method}}
\label{sec:format}

Fig.1 shows the overall architecture of our proposed model. Inspired by the GC ViT \cite{hatamizadeh2022global}, we construct a main autoencoder of the compression network based on an Aggregated-Window Transformer (AGWinT). The suggested entropy model is employed to efficiently encode the quantized latent representations into a stream of ones and zeros.

\subsection{\textbf{Encoder-Decoder}}




\begin{figure*}[tp]
    
  \centering
  \scalebox{0.61}{\includegraphics{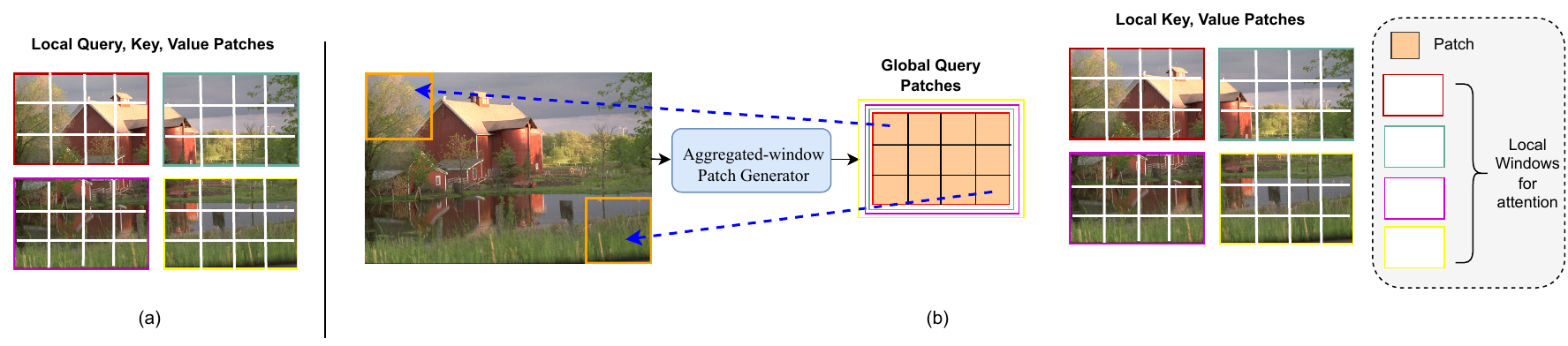}}
  \caption{ An illustration of the local and global attention mechanisms. (a) The local attention mechanism operates independently within each window, where each window possesses its specific query, key, and value patches. (b) The global attention mechanism extracts query patches from the entire input feature map, aggregating information from all windows. The global queries interact with local key and value patches, facilitating the capture of long-range information through cross-region interaction.}
  \label{fig:qplot}

\end{figure*}

\begin{figure}
  \centering
  \scalebox{.43}{\includegraphics{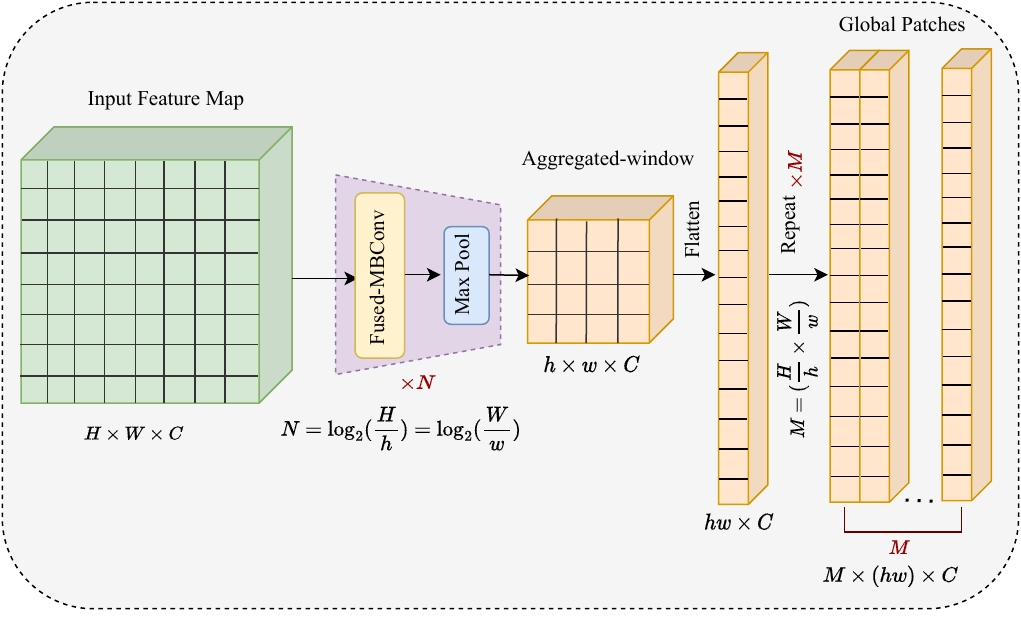}}
  \caption{ The diagram of the aggregated-window patch generator. The input feature map with dimensions $H$ and $W$ is fed into the module, which consists of a stack of $N$ identical layers. Each layer is composed of a Fused-MBConv block followed by a max pooling layer, which generates an aggregated window features. The number of layers $N$ is chosen such that the  extracted global feature dimension matches the size of the local window $h \times w$.  Subsequently, the global window is flattened and duplicated $M$ times, where $M=(\frac{H}{h} \times \frac{W}{w})$ corresponds to the number of local windows, to create the global query patches.}
  \label{fig:qplot}
\end{figure}

\subsubsection{\textbf{Encoder}}A hierarchical AGWinT is adopted as the encoder for obtaining feature representations at multiple resolutions by reducing spatial dimensions while increasing embedding dimensions by factors of two at 4 stages. Firstly, The input image ${x\in {R}^{
 H \times W \times3}}$ is fed into the patchify layer. which consists of a ${3\times 3}$ convolutional layer with a stride of 2 and the padding operation, to generate the overlapping patches. Then, the resulting patches are mapped into an embedding space with dimension C 
by using another ${3\times 3}$ convolutional layer. This layer is named embedding. To extract semantic features, each AGWinT stage is composed of multiple local and aggregated-window Transformer blocks. After  stages 1,2 and 3 a downsampling block is applied to  halve the spatial resolution of the feature map and double the channel number of the feature map.
\subsubsection{\textbf{Decoder}} An AGWinT decoder is the mirror of the encoder. We design the AGWinT decoder by replacing the patchify block with a unpatchify block, the embedding layer with a de-embedding layer, and the downsampling block with an upsampling block.
\subsubsection{\textbf{Downsampling block}}The role of this block is to generate hierarchical representation. The input feature map is passed through the Fused-MBConv \cite{tan2021efficientnetv2} to inject inductive bias into the network and model inter-channel correlations. Then, the convolution layer with a kernel size of 3 and stride of 2 is utilized to downsample the spatial feature resolution by 2, whereas the number of channels is doubled. The Fused-MBConv process can be written as: 
\begin{equation}
 \begin{aligned}
    &\hat{x}=DephWiseConv_{3*3}(x),\\ &\hat{x}=GELU(\hat{x}),
    \\&\hat{x}=SE(\hat{x}),\\
    &x=Conv_{1*1}(\hat{x})+x.
\end{aligned}
\end{equation}
Where SE \cite{hu2018squeeze} refers to the Squeeze and Excitation block and GELU represents the Gaussian Error Linear Unit function.

\subsubsection{\textbf{Local Transformer Block}}Like the Swin Transformer block \cite{liu2021swin}, the local Transformer block splits the input into local windows, then a window-based multi-head self-attention (W-MSA) computes the self-attention individually for each window. Computing local self-attention leads to extracting short-range information.
\subsubsection{\textbf{Aggregated-window Transformer Block}} As shown in Fig. 2, in contrast to the local Transformer block, where each local window has its specific query patches, the aggregated-window Transformer block employs global query patches to share across all windows to interact with local keys and values. Global query patches include information from the whole input feature map which is produced by an aggregated window patch generator once at every stage.
\subsubsection{\textbf{Aggregated-window Patch Generator}}This module consists of a stack of $N$ identical layers where each layer is composed of Fused-MBConv followed by a max pooling layer to extract global features, as displayed in Fig. 3. The number of layers is set to $N$ so that the output feature dimension matches the size of the local window. Then, the output is flattened and repeated to the number of windows to form the global query patches.


\subsection{ \textbf{Proposed Entropy model}}
As depicted in Fig. 4, the proposed entropy model is comprised of hyperprior-networks, local context, global context, and parameter network. The outputs of hyper-networks, local context, and global context are combined together and fed into the parameter network, comprising of $1\times 1$ convolution layers, to obtain the distribution parameters, i.e., $\mu$,$\sigma$. The process of deriving distribution parameters can be written as:
\begin{equation}
 \begin{aligned}
    & ({{\mu,\sigma}})=pm({\phi}_{sp},{\phi}_{ch},\psi_{l},\psi_{g}),\\ & {\phi}_{sp}= sh(\bm{y}), {\phi}_{ch}= ch(\bm{y}), \psi_{l}=lc(\bm{\hat{y}_{<i}}), \psi_{g}=gc(\bm{\hat{y}_{<i}}),
\end{aligned}
\end{equation}
\noindent where $pm(.)$, $sh(.)$, $ch(.)$ , $lc(.)$, and $gc(.)$ correspond to parameter network function, spatial-aware hyperprior model, channel-aware hyperprior model, local context, and global context, respectively. The local context is implemented using masked convolution with a kernel size of $5 \times 5$, as inspired by previous work \cite{minnen2018joint}. Detailed information about the other blocks will be presented in the following sections.\\


\subsubsection{\textbf{Global Context Block}}
This block is introduced to effectively exploit the global context information for obtaining a more accurate entropy model. It is based on an autoregressive model that uses all the previously decoded latents to predict the current decoding latent. As shown in Fig. 5(a), masked attention is used to model the global causal correlation. 

Due to the fact that the latent representations are serially decoded, the latents are unfolded into patches and then masked, represented as $p\in [H\times W, k \times k \times C]$ (where $H$, $W$, $k$, and $C$ correspond to height, width, unfold kernel size, and channels dimension, respectively) to calculate the masked attention. The normalized masked patches are considered as query tokens and key tokens (neglecting the magnitude effect of patches in calculating attention score), while regular masked patches are taken into account as value tokens. The formulation for the masked-multi-head attention can be expressed as follows:
\begin{equation}
 \begin{aligned}
    & Attention(\bm{q}, \bm{k}, \bm{v})= Concat(head_1,..., head_m)\bm{W},\\
    & head_i(\bm{q_i},\bm{k_i}, \bm{v_i})= softmax(\bm{q_i} \bm{{k_i}^T}\odot \bm{M})\bm{v_i},
\end{aligned}
\end{equation}
\noindent where $q_i ,k_i, v_i \in {R}^{
 HW \times \frac{K^2C}{h} }$ are the queries, keys, and values for the $i$-th head, respectively. $M\in {R}^{HW \times HW}$ represents a causal mask whose lower-triangular elements are one and the remaining elements are minus infinity.\\



  

%

\begin{figure}
    
  \centering
  \scalebox{.65}{\includegraphics{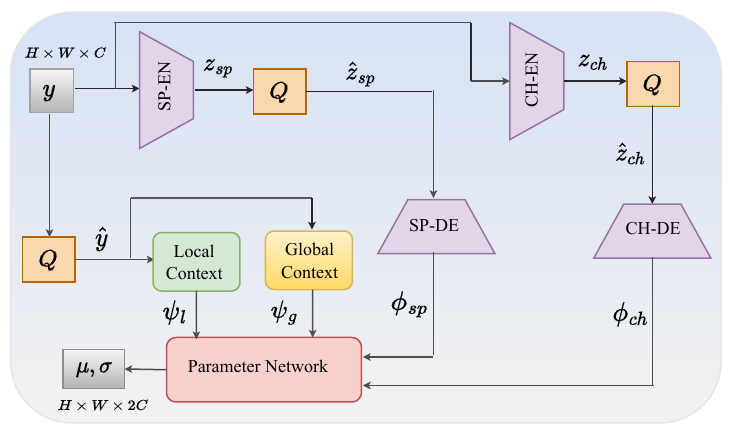}}
  \caption{Diagram of the proposed entropy model. Entropy coding operations for $z_{sp}$ and $z_{sp}$ are excluded for simplicity.}
  \label{fig:qplot}
\end{figure}

 \subsubsection{\textbf{Hperprior}}
 The newly proposed hyperprior is split into two distinct groups: the Channel-aware hyperprior $\bm{{\hat{z}}_{ch}}$ and the spatial-aware hyperprior $\bm{{\hat{z}}_{sp}}$. Each of these hyperpriors plays a crucial role in modeling different aspects of the dependencies within the latent representation $\bm{y}$. The Channel-aware hyperprior $\bm{{\hat{z}}_{ch}}$ is specifically designed to capture the inter-channel relationships at each spatial location of the latent representation, while the spatial-aware hyperprior $\bm{{\hat{z}}_{sp}}$ aims to model spatial dependencies within each channel of the latent representation. Both types of hyperprior are computed in parallel and complement each other, working together to jointly extract dependencies within the latent representation $\bm{y}$.

 \begin{figure}
    
  \centering
  \scalebox{.40}{\includegraphics{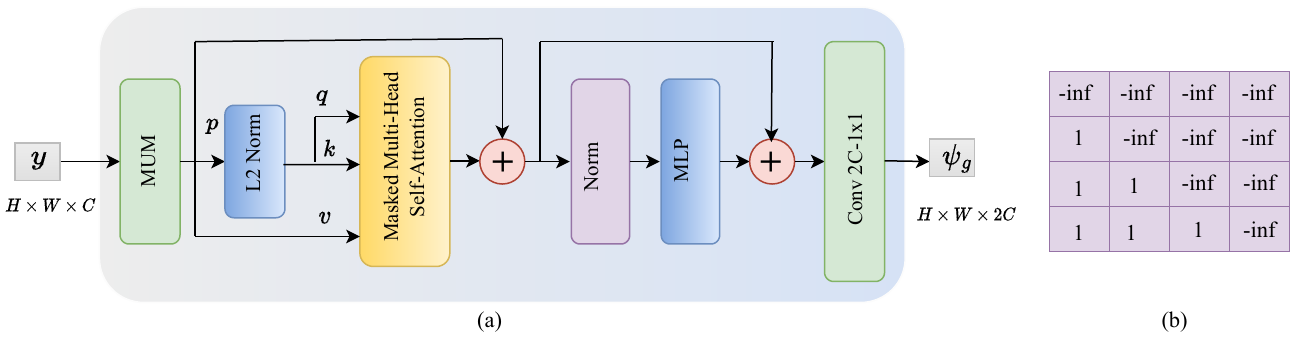}}
  \caption{Global Context Block: (a) The Transformer-based architecture, where $MUM$ stands for the masked unfolding module. (b) An example of causal mask $M$.}
  \label{fig:qplot}
\end{figure}



\textbf{Channel-aware Hyperprior:} We have developed the Channel-aware hyperprior model, which effectively captures inter-channel dependencies within the latent representation. As illustrated  in Fig. 6(a), this model is constructed based on an autoencoder architecture and utilizes channel-attention modules \cite{woo2018cbam} and stacks of $1\times1$ convolution layers to achieve its objective. The Encoder part generates the channel-aware hyperprior ${\bm{z_{ch}}\in {R}^{H \times W \times \frac{C}{16}}}$ from the input ${\bm{y} \in {R}^{H \times W \times C}}$, reducing the number of channels in the latent space while preserving spatial resolution. On the Decoder side, it takes the channel-aware hyperprior ${\bm{z_{ch}}}$ as input and generates the desired output ${\bm{\phi_{ch}}\in {R}^{H \times W \times 2C}}$.

\begin{figure}
    
  \centering
  \scalebox{.6}{\includegraphics{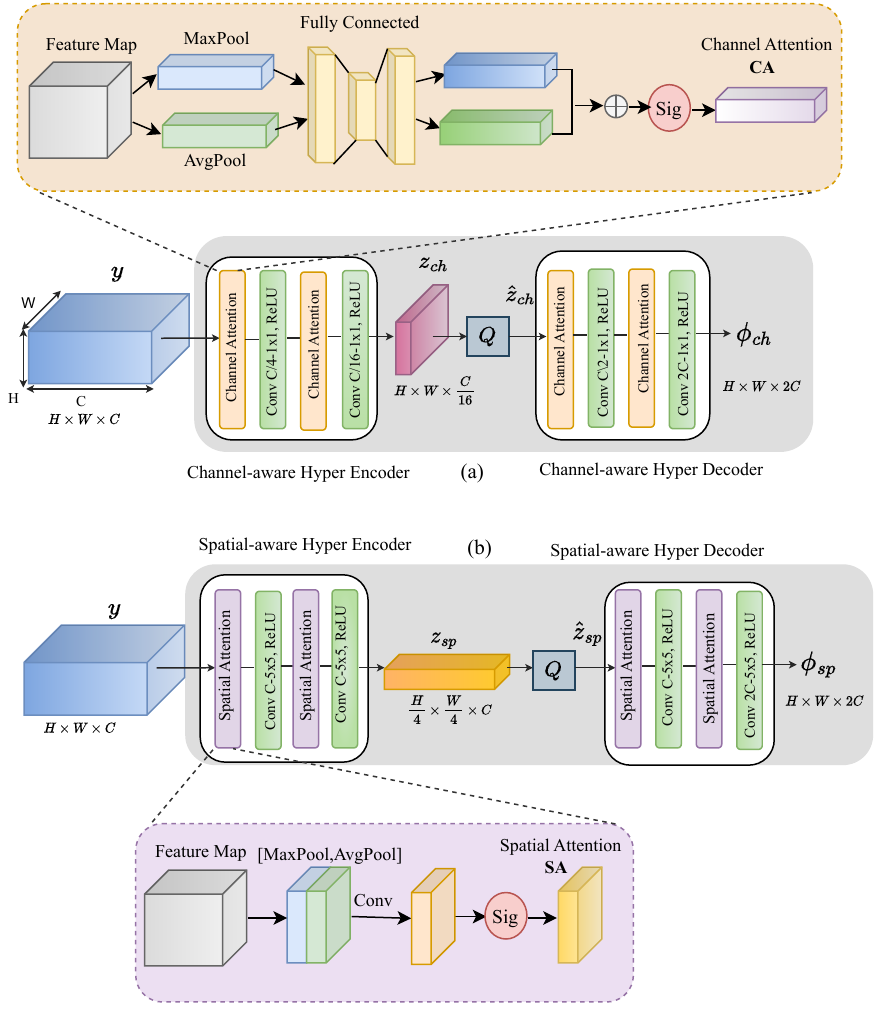}}
  \caption{ (a) The architecture of Channel-aware Hyperprior Network. (b) The architecture of Spatial-aware Hyperprior Network. Sig represents the sigmoid function.}
  \label{fig:qplot}
\end{figure}

\textbf{Spatial-aware Hyperprior:} Spatial-aware hyperprior model is designed to extract spatial interactions among the latent representation elements. The hyperprior ${\bm{z_{sp}}\in {R}^{\frac{H}{4} \times \frac{W}{4} \times C}}$, generated by utilizing this model, doesn't change the channel number of the latent representation; rather, it decreases the spatial resolution. The network adopted to output the spatial-aware hyperprior takes the form of an autoencoder architecture and consists of $5\times5$ convolution layers and spatial-attention \cite{woo2018cbam} blocks. As depicted in Fig. 6(b), the latent representation $\bm{y}$ is fed into an encoder block to obtain the spatial-aware hyperprior $\bm{z_{sp}}$, and the decoder component produces the output ${\bm{\phi_{sp}}\in {R}^{H \times W \times 2C}}$.\\

\subsection{\textbf{Quantization}} 
To facilitate end-to-end training feasibility, the quantization process requires a replacement with a soft differentiable function. In this study, we utilize uniform noise, which is added to the latent representations, as an approximation for the hard quantization operation \cite{balle2016end}. Consequently, the conditional probability of each latent, ${\hat{y}_i}$ is modeled as a univariate Gaussian with its location and scale convolved with a unit uniform distribution:
\begin{equation}
P_{\bm{\hat{y}} | \bm{\hat{z}}}(\bm{\hat{y}}|{\bm{\hat{z}},\bm{\theta}})=\prod_{i=1}(\mathcal{N}(\mu_{i},\,{\sigma_{i}}^2) \ast \mathcal{U}(-\frac{1}{2},\frac{1}{2}))(\hat{y}_i),
\end{equation}
 where location $\mu_{i}$ and scale $\sigma_{i}$ are determined by the entropy model, $\bm{\theta}$ denotes the parameters of the entropy model, $\bm{\hat z}=\{\bm{\hat{{z}}_{sp}}, \bm{\hat{{z}}_{ch}}\}$  is the quantized hyperpriors.

\subsection{\textbf{Loss Function}} 
Any learned image compression network attempts to jointly minimize the trade-off of rate and distortion, controlled by a Lagrangian multiplier which can be formulated as:
\begin{equation}
    R+\lambda D
    \label{eq:rate-distortion},
\end{equation}
where $R$ denotes the estimated rate of the quantized latent representation and $D$ is the distortion between the input image and reconstructed image. Since the probability distribution of the latent representation is conditionally estimated by hyperpriors, the rate term $R$ consists of the estimated entropy of the latent representation and two hyperpriors. Therefore, the rate term can be described as below:
\begin{equation}
\bm{E}_{x\sim p_X}[-\log_2P_{\bm{\hat{y}}|\bm{\hat{z}}}(\bm{\hat{y}}|\bm{\hat{z}})-\log_2P_{\bm {\hat{z}_{ sp}}}({\bm{\hat{z}}_{sp}}))-\log_2P_{\bm {\hat{z}_{ch}}}(\bm{{\hat{z}}_{ch}}))],
\end{equation}
\noindent where $\bm{\hat z}=\{\bm{\hat{{z}}_{sp}},\bm{\hat{{z}}_{ch}}\}$ includes both hyperpriors and the non-parametric fully factorized entropy model \cite{balle2017endtoend} is employed to approximate the probability distribution of hyperpriors.


\section{\textbf{EXPERIMENTS}}
\subsection{\textbf{Implementation Details}}
For our experiments, we utilize a combined set of $5,285$ high-resolution images from datasets DIV2K, Flickr2K, and CLIC2021 \cite{clic} as our training dataset. The evaluation set consists of images from the Kodak dataset \cite{kodak}. During the training phase, we use a batch size of $16$, which includes randomly cropped $256\times256$ patches. To cover a wide range of bitrates, we set the hyper-parameter $\lambda$ to take values from the set $\{0.00125, 0.005, 0.01, 0.02, 0.04, 0.08\}$. All the models are trained for 100 epochs using the Adam optimizer \cite{kingma15adam} for a total of $2.4$ million steps. The initial learning rate is set to $10^{-4}$ and gradually decreased to $1.2\times10^{-6}$ during the training process.


\begin{figure}
    
  \centering
  \scalebox{.4}{\includegraphics{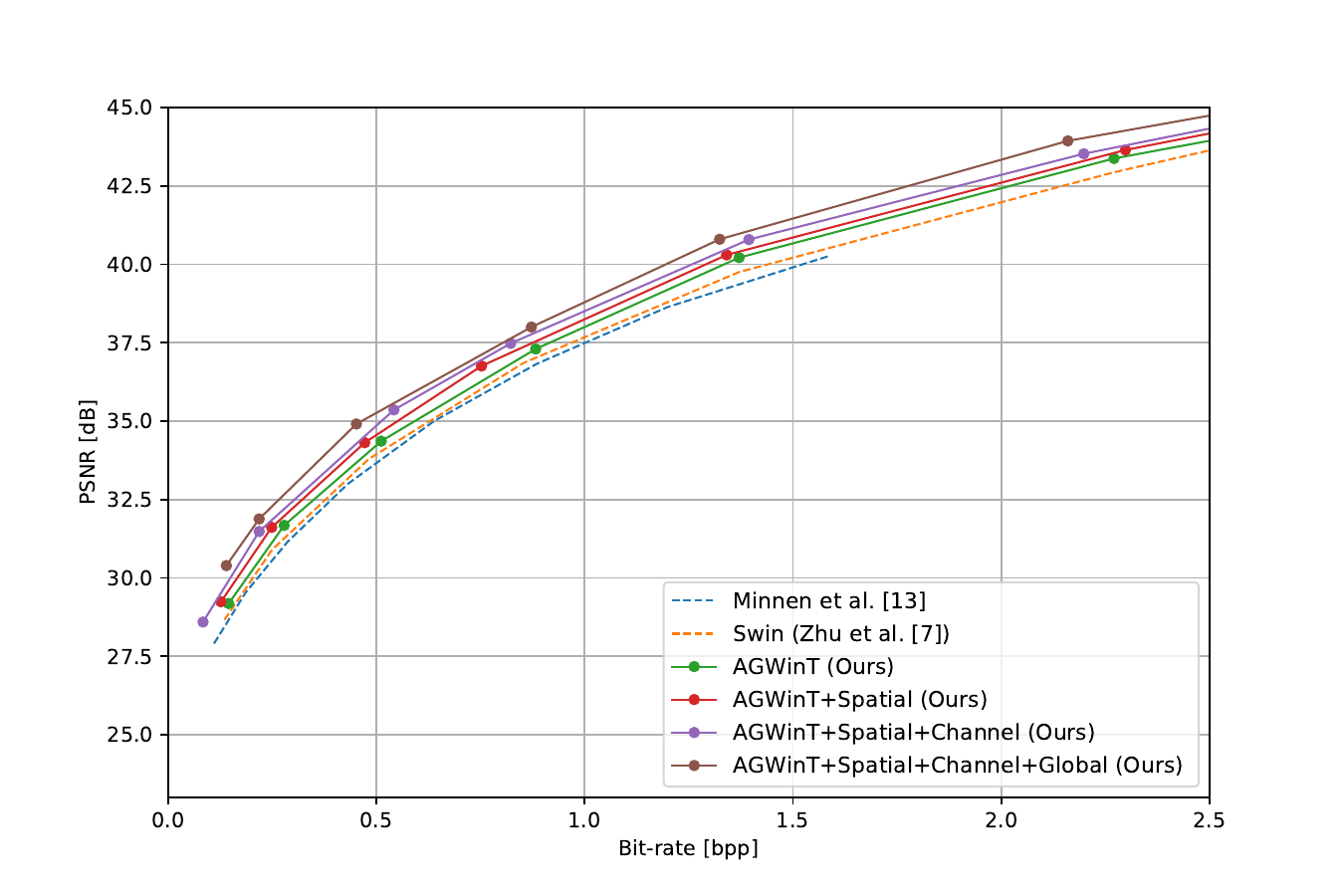}}
  \caption{Rate-distortion performance comparison of different methods on the Kodak dataset, comprising 24 images. Distortion is measured by PSNR.}
  \label{fig:qplot}
\end{figure}

\begin{figure}
    
  \centering
  \scalebox{.4}{\includegraphics{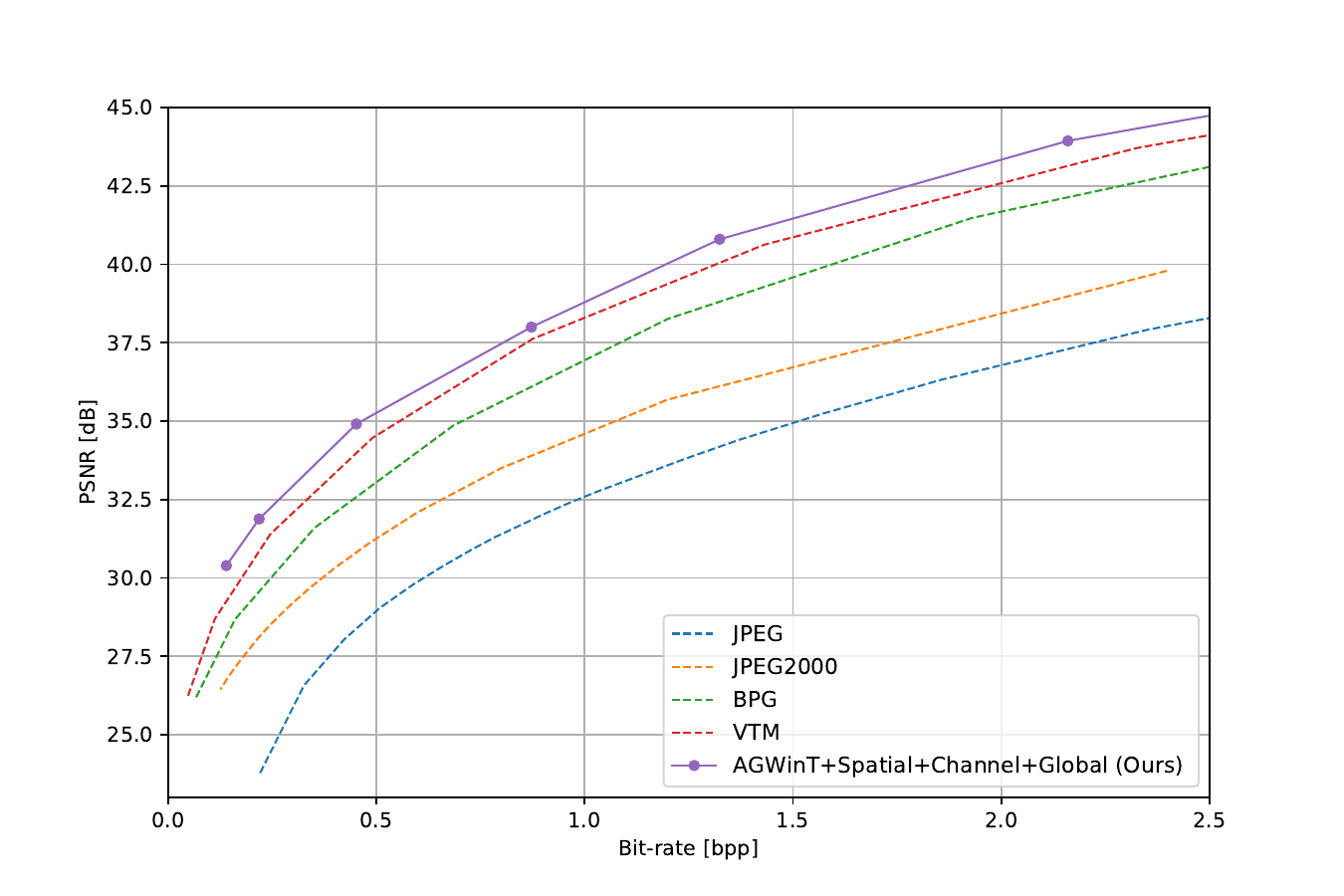}}
  \caption{ Comparison of our full model with traditional image codecs in terms of rate-distorsion performance. The RD values are averaged over 24 images from the Kodak test set. }
  \label{fig:qplot}
\end{figure}

\begin{figure*}[tp]
    \centering
    \includegraphics[width=.83\linewidth]{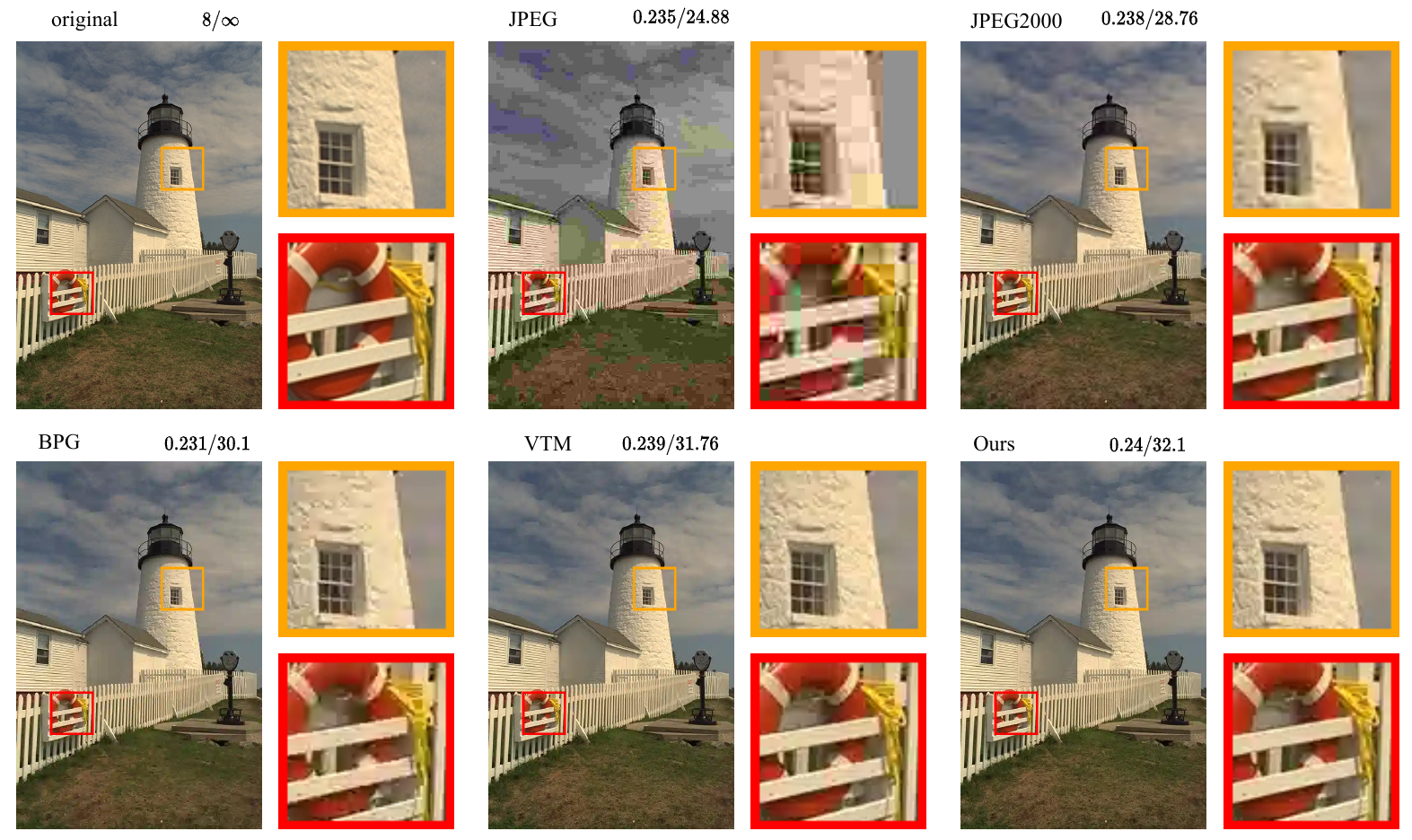}
    \caption{Visual comparison of our proposed framework with other conventional image codecs, using the bit-rate/distortion [bpp$\downarrow$/PSNR$\uparrow$] as the metric. The results demonstrate that our method achieves lower distortion in terms of PSNR compared to the other codecs, highlighting its superior capability in preserving image quality.  }
    \label{fig:visual-comparison}
\end{figure*}
\subsection{\textbf{Ablation Study}}
In this section, we conduct several ablation studies to investigate the effectiveness of our proposed Transformer-based autoencoder and the entropy model. In the first study, to assess the performance of AGWinT, we choose the Minnen \emph{et al.} model \cite{minnen2018joint}, whose entropy model is comprised of hyperprior and local context, and replace the main convolutional autoencoder with the Swin Transformer-based autoencoder and AGWinT Transformer-based autoencoder.  The results demonstrate that the AGWinT-based network outperforms both the convolutional and Swin-based architectures in terms of rate-distortion performance. This improvement can be attributed to the fact that AGWinT is capable of effectively capturing long-distance and local correlations, leading to enhanced compression performance. 

In the second study, to validate the impact of decomposing hyperprior and adding spatial and channel attention, we build three different hyperprior models: existing hyperprior, the whole compression model is referred to as the "AGWinT" framework, "spatial-aware hyperprior", and "spatial-aware hyperprior + channel-aware hyperprior". Each of the mentioned hyperprior frameworks is combined with local context to form the entropy model. These entropy models are then integrated into compression models based on the AGWinT autoencoder. As depicted in Fig. 7, the "spatial-aware hyperprior + channel-aware hyperprior" model exhibits better performance compared to the other models. This enhancement is related to the improved modeling capabilities of the combined spatial and channel attention in exploiting the complex dependencies within the latent representations.

In the third study, we examine the effect of incorporating global context in the entropy model. Fig. 7 demonstrates that global context contributes to achieving a better result. This evaluation emphasizes that combining global context with local context leads to precise modeling correlations of the latent representation.    

 \subsection{\textbf{Comparison with Traditional Codecs}}
 We compare the rate-distortion (RD) performance of our full model, denoted as "AGWinT+Spatial+Channel+Global", with traditional video compression standards, such as JPEG \cite{jpeg}, JPEG2000 \cite{jpeg2000}, HEVC-Intra (BPG) \cite{bellard2015bpg}, and VVC-Intra (VTM) \cite{vtm2022}.  The distortion
is measured by the Peak Signal-to-Noise Ratio (PSNR). 
Fig. 8 illustrates that our proposed model outperforms the JPEG, JPEG2000, and BPG codecs in terms of RD performance. Additionally, it showed comparable performance to VTM. 

\subsection{\textbf{Visual Quality}}
 In Fig. 9, we present an example of a reconstructed image (kodim19.png) obtained using our proposed framework, as well as the compression standards JEPG \cite{wallace1992jpeg}, JEPG2000, BPG, and VTM. The visual comparison reveals that our reconstructed image retains significantly more details while achieving a similar bpp value. This observation highlights the superior performance of our approach in preserving image quality.

\section{\textbf{Conclusion}}
\label{sec:pagestyle}

This paper presents a novel image compression algorithm using a Transformer-based autoencoder and a new entropy model. The main autoencoder consists of a local Transformer and aggregated-window Transformer blocks which are responsible to capture both long-term and short-term relationships for achieving a more compressible and decorrelated representation of the input image. Our proposed entropy model is comprised of two different hyperpriors: channel-aware hyperprior and spatial-aware hyperprior. 
These two hyperpriors aim to model the remaining spatial and cross-channel redundancies in the latent representation. The global context is integrated with local context and hyperpriors to exploit global information which assists in accurately estimating the distribution of the latent representation. Our experimental results demonstrate that our proposed network boosts compression efficiency.

\bibliographystyle{IEEEtran}
\bibliography{mybib}
\end{document}